\documentclass[a4paper,11pt]{article}
\pdfoutput=1

\usepackage[english]{babel}
\usepackage[utf8x]{inputenc}
\usepackage[T1]{fontenc}
\usepackage{ upgreek }
\usepackage{gensymb} 
\makeatletter
\newcommand{\ssymbol}[1]{^{\@fnsymbol{#1}}}
\makeatother

\usepackage{amsmath}
\usepackage{graphicx}
\usepackage[colorinlistoftodos]{todonotes}
\usepackage[colorlinks=true, allcolors=blue]{hyperref}

\usepackage[a4paper,top=3cm,bottom=2cm,left=3cm,right=3cm,marginparwidth=1.75cm]{geometry}

\usepackage{authblk}
\title{The Smeared Null Energy Condition}
\author[$\ssymbol{2}$]{Ben Freivogel}
\author[$\ssymbol{2}$]{Dimitrios Krommydas}
\affil[$\ssymbol{2}$]{ ITFA and GRAPPA, Universiteit van Amsterdam,
Science Park 904, Amsterdam, the Netherlands}

\begin{document}
  \maketitle
  
\large

\begin{abstract}
\large
We propose a new bound on a weighted average of the null energy along a finite portion of a null geodesic: the Smeared Null Energy Condition (SNEC). We believe our bound is valid on scales small compared to the radius of curvature in any quantum field theory that is consistently coupled to gravity. If correct, our bound implies that regions of negative energy density are never strongly gravitating, and that isolated regions of negative energy are forbidden.
\normalsize
\end{abstract}

\newpage

\tableofcontents







\medskip

\section{Introduction and Outline}
\label{sec:intro}

General Relativity 
leaves an important question unanswered: Which metrics correspond to physical spacetimes? To  resolve this issue, restrictions on the allowed form of the stress energy tensor are needed. Various energy conditions have been proposed that exclude worrisome spacetimes like warp drives, traversable wormholes, perpetual motion machines, etc., which lead to causality paradoxes.



The Null Energy Condition (NEC) is the most important of these energy conditions. It states that
\begin{equation}
 T_{\mu\nu} K^\mu K^\nu  \geq 0 ,
\end{equation}
where  $T_{\mu\nu} $ is the stress energy tensor and $K^{\mu}$ is any null vector. The NEC is believed to hold for all sensible classical matter. In addition, the NEC is sufficient to prove the absence of traversable wormholes, the area theorem, and the focusing theorem.




 
An obvious generalization of the NEC to Quantum Field Theory (QFT) is:
\begin{equation}
\label{naive qnec}
\langle \psi | T_{\mu \nu}(x^\alpha) K^\mu K^\nu | \psi \rangle \geq 0 ~,
\end{equation}
for all states $|\psi \rangle$ and all spacetime points $x^\alpha$. If true, this condition would allow us to extend the theorems relying on the NEC to the semiclassical regime, where classical gravity is coupled to the expectation value of the stress tensor,
\begin{equation}
\label{sca}
G_{\mu \nu} = 8 \pi G_N \langle T_{\mu \nu} \rangle ~.
\end{equation}

However, there exist states in even the simplest QFT's that violate condition \eqref{naive qnec}  \cite{Krommydas:2018fgs}, leaving the question of which spacetimes are physical unclear. Interesting recent work has proposed and/or proved quantum generalizations of the NEC that are true in QFT. Two important examples are the Averaged Null Energy Condition (ANEC) \cite{Borde:1987qr}-\cite{Faulkner:2016mzt}, which constrains the integrated null energy along a complete null geodesic to be positive definite, and the Quantum Null Energy Condition (QNEC) \cite{Bousso:2015mna}-\cite{Fu:2017evt}, which relates the negativity of the null energy to the entanglement of the quantum fields.

Despite these beautiful results, we would like to have something more: we want a local constraint on NEC violation that does not depend on additional quantities like the entanglement entropy. Such a local constraint would be very useful in diagnosing, for example, what types of traversable wormholes are possible.  Traversable wormholes connecting distant points in the same spacetime are not excluded by the ANEC. 

Negative values for the null components of the stress tensor are  problematic when they backreact on the metric. Therefore, NEC violation is important in the semiclassical approximation (\ref{sca}). Clearly, this equation is only sensible if the fluctuations in the stress tensor are small compared to its mean, so that it makes sense to replace the stress tensor by its expectation value.

The stress tensor at a point has infinite 2-point function, so the fluctuations are never small. However, the average of the stress tensor over some region, which we call the `smeared' stress tensor following Ford and collaborators, can have finite fluctuations for appropriate regions. The smeared stress tensor is characterized by the smearing length $\tau$; we define it more carefully below. It is plausible that the semiclassical approximation is good as long as the smeared stress tensor has small fluctuations for some smearing length small compared to the curvature scale of the geometry.



\paragraph{False Conjecture.} By looking at a large number of examples, we initially wanted to propose the following conjecture: {\bf The smeared stress tensor satisfies the NEC whenever its fluctuations are small compared to its mean.} 

Although this conjecture is simple and passes a number of tests, it is sadly not true in general. We argue below that this conjecture is violated in theories with a large number $N$ of weakly interacting fields. Increasing the number of fields makes the fluctuations small compared to the mean as $1/\sqrt{N}$, as is familiar from statistics.

In addition, there is a powerful argument against this conjecture even in theories with a small number of fields.\footnote{We thank Ken Olum for explaining this argument to us.} Given any state with negative expectation value of the smeared stress tensor, there must exist eigenstates of the smeared stress    tensor with negative eigenvalues, and these have zero fluctuations.


It may be that some refinement of this false conjecture is still correct, at least in theories with a reasonable number of fields, as we show through a number of examples in section \ref{our SCG stuff}.



\bigskip

\paragraph{Proposed Bound.}
There are no existing  bounds on the average of the stress tensor along a finite portion of a null geodesic above $1+1$ dimensions. 
Our best guess is the new bound:   

\begin{equation}
\label{Bound intro}
\boxed{\langle T^s_{kk} \rangle \geq - \frac{B}{G_N \tau^2}~.} 
\end{equation}
Here $T_{kk}$ is the null component of the stress tensor, smeared over an affine distance $\tau$ along a null geodesic. We propose that this formula is true in any region where perturbative quantum gravity (defined more carefully below) is a good approximation, under the condition that the smearing length $\tau$ is small compared to the curvature scale of the geometry.  

We have not determined the constant $B$, but our proposal is that a single, order one number can be chosen such that all consistent theories satisfy the bound. We define the averaging procedure more carefully below.

The reason that this bound may hold for an arbitrary number of fields is that the stress tensor scales linearly in the number fields, but the Newton constant scales inversely with the number of fields, $G_N \sim 1/N$ \cite{Kaloper:2015jcz} for a large number of fields.

The quantity that is bounded by this conjecture is related by the Raychadauri equation to the focusing of null geodesics. As we motivate in more detail below, this conjecture claims that whenever geodesics `de-focus' due to negative null energy, they do so by a small amount that remains in the linear gravity regime.

Roughly, our bound prohibits isolated regions of negative null energy in a way that is qualitatively similar to the bounds of Ford and collaborators, with the novel feature that it applies to null, rather than timelike, smearing. To see that it prohibits isolated regions of negative energy, suppose that there is a localized region of NEC violation, with no nearby positive energy. The smeared stress tensor for a region including the NEC violation would be
\begin{equation}
\langle T^s_{kk}  \rangle \propto - \frac{b}{\tau} ~,
\end{equation}
where $\tau$ is the smearing length and $b$ is a constant (independent of $\tau$) amount of negative energy per unit transverse area. For large $\tau$, this will violate our proposed bound (\ref{Bound intro}), which falls off as $\tau^{-2}$. Therefore, our bound prohibits isolated regions of NEC violation.

We claim that our bound is valid in the regime well-described by perturbative quantum gravity. By perturbative quantum gravity, we mean a situation where the spacetime is well-described by a classical background plus perturbative metric fluctuations. Specifically, the stress tensor is split into a `classical' c-number piece (most commonly, but not necessarily, the expectation value) and a fluctuation piece. The metric is split into a classical and quantum part. The classical part of the metric solves the Einstein equation, with the classical stress tensor as source, while the quantum fluctuations are coupled using the framework of quantum field theory in curved spacetime. With this definition, perturbative quantum gravity has a wider regime of applicability than semiclassical gravity, where the metric is treated classically.

Our bound is naturally interpreted geometrically. Moving the factor of $G_N$ to the other side gives, via Einstein's equations, null components of the Ricci tensor. Our bound then takes the form
\begin{equation}
R^s_{kk} \geq - \frac{\#}{\tau^2}
\end{equation}
Note that this form of the bound does not have any factors of $\hbar$, so it makes sense purely classically.\footnote{We thank the referee for bringing the issue of the $\hbar$ dependence to our attention.}

On the other hand, one might like to interpret the bound in field theory language. Here one would hope to see that NEC violation is not possible in the classical limit.  In field theory factor of $G_N$ is not natural, but a UV cutoff scale can appear. Writing $G_N = \hbar l_P^{D-2}$ and keeping $\hbar$ explicit, our bound becomes
\begin{equation}
    T^s_{kk} \geq - \frac{B \hbar}{l_P^{D-2} \tau^2} ~.
\end{equation}
For a field theory with $N$ fields, we can use the lore \cite{Kaloper:2015jcz}
$
G_N \leq l_{UV}^{D-2}/N~,
$
where $l_{UV}$ is the ultraviolet cutoff of the theory, to obtain
\begin{equation}
    T^s_{kk} \geq - \frac{B \hbar N}{l_{UV}^{D-2} \tau^2}
\end{equation}
This bound now makes sense in field theory and has the expected behavior that NEC violation goes away in the classical limit.

There is one annoying aspect of our proposal. The operator we focus on is the average of the stress tensor over a single null geodesic. However, we will show that this operator has infinite fluctuations in the vacuum in quantum field theory above 1+1 dimensions. We show (in free field theory) that additional smearing over the orthogonal null direction is needed in order to tame these divergences. 

Nevertheless, a bound on the expectation value of our operator is sensible;  other proposed quantum versions of the NEC also bound operators with infinite fluctuations. It is an interesting problem for future work to smear over additional directions to obtain an operator with finite fluctuations, and we believe our bound on the expectation value will extend to these operators; one may also expect that further smearing will allow for a stronger bound. Our focus here is on theories that are coupled to gravity, so we do not pursue a bound on an operator with finite fluctuations within field theory.


It is plausible that one could use our bound to prove generalized versions of Penrose-Hawking-type singularity theorems, using a procedure similar to the one used by Fewster in \cite{Fewster:2010gm}. 
To our knowledge, there is no obvious relation between our bound and the Generalized Second Law, QFC, QNEC or any other entropic type of bound. In a sense, this is one of the most intriguing parts of our proposal since it allows one to impose restrictions without necessarily calculating entropies and their derivatives, whose computation is often a challenging enterprise. That said, it would be very interesting if an explicit connection between our proposal and these entropic bounds were to be established. 

\bigskip

The paper is organized as follows. In chapter \ref{previous work}, we summarize a collection of the most relevant previous work: the global ANEC \cite{Borde:1987qr}-\cite{Faulkner:2016mzt}, the semi-local Quantum Inequalities (QI's) \cite{Ford:1978qya} and finally the local QNEC \cite{Bousso:2015wca}. 
In chapter  \ref{our SCG stuff}, 
we present the results for the expectation value of the stress-tensor operator in some novel and already studied NEC violating states \ref{interesting states}. In section \ref{conjecture} we present our conjecture, 
along with several supporting arguments. In theories where it is obeyed, our conjecture provides the only known way to exclude all exotic space-times, and generalize all GR results to the Semi-Classical Gravity (SCG) regime. 
In section \ref{New Bound} we discuss our proposed bound \eqref{Bound}, which we then motivate with a number of gedanken experiments. In \ref{explicit eg} we show how the explicit example of the vacuum plus two particles state satisfies bound \eqref{Bound}. Finally, we calculate the fluctuations of a number of smeared stress energy operators.

\section{Previous Work}
\label{previous work}

In this chapter we briefly discuss the most prominent of the already existing bounds on Quantum NEC violating matter. Bounds on exotic matter can be categorized as global, local, or semi-local. Each category has some general strengths and shortcomings which we discuss throughout this chapter.

\subsection{Global, Local and Semi-Local constraints}
\label{global and local bounds}

\medskip

\textbf{ {Averaged Null Energy Condition (ANEC):}} Historically the first to suggest a global energy condition was Tipler in 1977 \cite{Tipler:1978zz}. Tipler considered the average over one of the point-wise energy conditions, namely the Weak Energy Condition. The averaging was taken over the whole worldline of the observer, and the resulting bound is known as the Averaged Weak Energy Condition. This condition's counterpart for null geodesics, goes by the name of Averaged Null Energy Condition:

\begin{equation}
\int^{\infty}_{-\infty} \langle T_{\mu\nu} K^\mu K^\nu \rangle d\lambda \geq 0 ~.
\end{equation}

The statement of the bound, is that the integral over the entirety of a null geodesic with affine parameter $\lambda$ of the projected $T_{\mu\nu}$ on a null vector $K^{\mu}$ (tangent to the geodesic), must always be greater or equal than zero.

This weaker condition has been utilized in order to salvage certain GR results in spacetimes with (quamtum) NEC violating matter content \cite{Borde:1987qr}. 
In 1991, Wald and Yurtsever \cite{Wald:1991xn} proved that ANEC holds in two and four-dimensional Minkowski spacetimes, for any Hadamard state along achronal null geodesics. Since then there have been significant advances in the field, such as the work of Kontou and Olum \cite{Kontou:2015bta} who proved the achronal ANEC on curved backgrounds obeying NEC, using an entirely different method. As Olum \textit{et al} showed in \cite{Graham:2007va}, this achronal ANEC can be used to exclude closed timelike curves and certain wormhole solutions. Recently, Hartman {\it et al.} \cite{Hartman:2016lgu} and Faulkner {\it et al.} \cite{Faulkner:2016mzt} have used information theoretic quantities to prove ANEC in several interesting set-ups.



As we discuss in section \ref{our SCG stuff}, there exist NEC violating states which when integrated lose their negative contributions, making them good candidates for the construction of wormhole solutions which do not violate ANEC. Wormhole spacetimes with arbitrarily small or no ANEC violations are discussed in \cite{Rahaman:2007qd}- \cite{Jamil:2009vn}. Some of the most important results of GR, are the Penrose-Hawking singularity theorems \cite{Senovilla:2014gza}. Unfortunately, ANEC and other global averages are not always sufficient to replace local condition as postulates for their proof. Finally, ANEC can be used to exclude certain wormholes that connect asymptotically flat regions of ``different'' universes. It cannot be used to exclude wormholes in the same universe. 

\textbf{{Quantum Inequalities (QI's)}} are a set of semi-local bounds in QFT, originally introduced by Ford in 1978 \cite{Ford:1978qya}. The quantum inequalities are more local than the averaged energy conditions since they involve integrals over only a specific interval of interest and not an entire direction. They are essentially restrictions on the magnitude/time duration of the negative energy density (or flux) in some time interval. The form of these restrictions is the following:

\begin{equation}
\label{QI}
\langle \tilde{\rho} \rangle =\int^{\infty}_{-\infty} \langle \rho \rangle f(\tau) d\tau  \geq  -\frac{C\hbar}{c^3 \tau^{4}_0}\;.
\end{equation}




Here $\tau$ is the proper time measured by the observer and $f(\tau)$ is a sampling function (a smooth peaked function of width $\tau_0$, which integrates to 1). Equation \eqref{QI} tells us that the amount of the smeared negative energy density $\hat{\rho}$ is inversely proportional to the fourth power of the duration of the interval $\tau^4_0$. Simply put, the more negative energy one sees, the shorter it can exist. We should note that QI's resemble the familiar uncertainty principle, although the later has not been assumed for their derivation.




These constraints have been proven for a massless and massive free scalar, E/M fields, a massive spin-1 field and a Dirac field, for arbitrary smooth sampling functions in 2-D \& 4-D Minkowski and in a few curved spacetimes \cite{Fewster:2012yh}. Recently, QI's were proven for all 1+1-CFT's with causal Holographic duals \cite{Levine:2016bpj}.

Going back to the traversable wormhole example we note that from the bounds discussed so far, QI's place the strongest restriction on their existence \cite{Fewster:2005gp}. If one wishes to have a sufficiently big, macroscopic traversable wormhole (so that he can travel through it comfortably), he would in general need a huge amount of negative energy. This, in turn, implies that the wormhole would remain open for a ridiculously small amount of time. 
Finally, QI's have been used to replace NEC for the proof of Penrose-Hawking-type singularity theorems \cite{Fewster:2010gm}.

 \textbf{Quantum Null Energy Condition:} One of the most exciting recent developments which reignited the interest in the field of Energy Conditions is the Quantum Null Energy Condition (QNEC) \cite{Bousso:2015wca}, introduced by R. Bousso, A. Wall \textit{et al.} \cite{Bousso:2015mna}. The original goal of \cite{Bousso:2015mna} was to unify the Covariant Bousso bound \cite{Bousso:1999xy}  with the classical focusing theorem, using a universal inequality conceived by the same authors; the Quantum Focusing Conjecture (QFC). The Quantum Focusing Conjecture is a proposal for a quantum corrected version of the Classical Focusing Theorem. The quantum null energy condition arises naturally by imposing this conjecture on a specific setup.

 In a nutshell, QNEC is the following statement: for an entangling surface $\Sigma$ with a normal, null hypersurface $N$ the expectation value of the null-null projected stress-energy operator on $N$, is bounded by the second order deformation of the geometric entanglement entropy outside the entangling surface, along $N$.\footnote{Note that although the entanglement entropy contains non-local information, the object appearing in this bound is local.} Our goal in this work is complementary to the QNEC proposal: we propose a c-number bound on the smeared stress tensor, while QNEC bounds the local expectation value of the stress tensor in terms of derivatives of the entanglement entropy, which depends on the quantum state.

QNEC has been proven for free and super-renormalizable bosonic field theories \cite{Bousso:2015wca} and for Holographic theories formulated in flat spacetime \cite{Koeller:2015qmn}. It was later shown by D. Marolf \textit{et al}. \cite{Fu:2017evt} that QNEC can be proven for certain Holographic theories on arbitrary backgrounds. QNEC and QFC were analyzed on curved backgrounds in \cite{Akers:2017ttv}. In \cite{Fu:2017evt}, one can find a detailed analysis of the assumptions required for QNEC's proof and finiteness in different dimensions. Recently, T. Faulkner and collaborators \cite{Balakrishnan:2017bjg}, proved a stronger, more general version of QNEC. 

\textbf{ Connection of NEC to thermodynamics and strings.} 
J. P. van der Schaar and M. Parikh showed that the NEC can be related to the imposition of the Virasoro constraint on the world-sheet of a bosonic string \cite{Parikh:2014mja}. In later work, M. Parikh \textit{et al.} provided a connection between the NEC, the second law of thermodynamics and the Bekenstein entropy \cite{Parikh:2015wae}. 

\medskip

\section{On Violations of the Null Energy Condition and the Semi-Classical Gravity approximation}
\label{our SCG stuff}

We start this chapter by calculating the expectation value of the stress tensor in certain classes of states which we find noteworthy. We then proceed to motivate and state our conjecture, and finally test it for a number of representative classes of states. The computations are performed in flat background, which is a good first order test of our conjecture. Our findings support the validity of our proposal for theories with a reasonably small number of fields.

\subsection{The stress tensor for some interesting states}
\label{interesting states}

The first class of states we consider are NEC violating superpositions of eigenstates of the number operator. The negative energy density in these states arises in the form of oscillatory terms. Thus, when integrated over an entire Cauchy slice only the positive term remains. One could be tempted to use matter in this states to construct traversable wormholes, but as we shall see in \ref{tests} these states are too quantum for that.

\textbf{Oscillating states}. Let us consider a superposition of the vacuum and a two-particle state with a \textit{specified} momentum k. The calculation of the expectation value of the stress tensor in this state was originally performed by L.H.Ford and M.J. Pfenning \cite{Pfenning:1998ua}.   



The result for the expected value of the energy density in this state reads:
\begin{equation}
\label{0+2 SET}
 \langle \psi \rvert : T_{tt} : \lvert \psi \rangle_{0+2} =  \frac{w}{2L^3}\left[1- \frac{\sqrt{6}}{2} \cos(\textbf{kx}-wt) \right]~.
\end{equation}
Here $: T_{tt} :$ is the usual normal ordered (vacuum subtracted) energy density component of the stress tensor operator.

Note that such states can also be found in the spectrum of 1+1 CFT's. For example take a superposition of a primary field $\mathcal{O}_{1}\lvert 0 \rangle$ 
of conformal weight $h_1$ and its first descendant  $\mathcal{O}_{2}\lvert 0 \rangle = L_{-1} \mathcal{O}_{1}\lvert 0 \rangle$ (where $L_{-1}$ is the usual raising operator). A similar oscillatory term appears in the energy density, which now also depends of the conformal weight. For certain reasonable values of the conformal weight the energy density is again negative in the same way as in (\ref{0+2 SET}). 

\textbf{Perturbed Vacuum states.} We saw above that  it is relatively easy to find NEC violating states even in 1+1 CFT's. However, consider the class of states that are obtained by acting on the vacuum with a source,
\begin{equation}
\label{pert vac}
e^{-i\int d^{2}x J(x) \Phi(x)}\lvert 0 \rangle ,
\end{equation}
where $J(x)$ is a small source, and $\Phi(x)$ a free massless scalar field. 
Computing the expectation value of the null-null stress tensor,

\begin{equation}
\label{pert stress}
\begin{split}
\langle T_{++} \rangle_{per.} & = \langle 0 \rvert e^{-i\int J (x) d^{2}x\Phi (x)} T_{++} e^{i\int J (x) d^2x\Phi (x)}\lvert 0 \rangle \\
& = \langle 0 \rvert \left(1 -i\int d^2x J (x)\Phi (x)-\frac{1}{2}\int d^2x d^2w J (x) J(w)\Phi (x)\Phi (w) \right) \cdot \left( \partial_{+}\Phi(z)\partial_{+}\Phi(z) \right) \cdot \\ 
& ~~~~ \cdot \left(1 + i\int d^2w J(w)\Phi(w)-\frac{1}{2}\int d^2x d^2w J (x)J(w)\Phi (x)\Phi(w) \right)  \lvert 0 \rangle
 = \\ & =  \langle T_{++} \rangle_0 + \frac{1}{4\pi}   \left( \int dx_-  J(z_+,x_-)  \right)^2 .
\end{split} 
\end{equation}


The full calculation can be found in Appendix \ref{Perturbed vacuum states}. This result shows that there is a general class of states for which the quantity of interest is positive definite. We take this as evidence that the spectrum of NEC violating states in QFT's is smaller than one would generally expect.

\subsection{Conjecture}
\label{conjecture}



Our primary interest in bounds on the behavior of the stress tensor is in constraining physically allowed metrics through Einstein's equation. The connection to the metric is most direct when we can use the semi-classical approximation, coupling the classical Einstein equations to the expectation value of the stress tensor:
\begin{equation}
G_{\mu\nu}= 8\pi G_{N} \langle T_{\mu\nu} \rangle_{\psi}~.
\end{equation}
As we have already mentioned, there exist states in the spectrum of even the simplest of QFT's which violate NEC. However, not all QFT states lead to expectation values which correspond to reasonable macroscopic observables. In order for this equation to make sense, the probability distribution function(al) of the eigenvalues of the (smeared) stress-energy operator must be sharply peaked around its mean (expectation value of the stress-energy operator in that state). Inspired by the above, we propose the following conjecture:

 



\paragraph{False Conjecture.} {\bf For states in which $\langle T^s_{kk} \rangle_\psi < 0$, SCG is not a valid approximation.}

\medskip

The SCG approximation is valid as long as several conditions are met. The fist condition is that the curvature scales are far from the Planck scale, where we expect quantum gravity to kick in. Another condition that has received less attention, but has been studied nonetheless \cite{Kuo:1993if}, is that the fluctuations of the stress tensor must be much smaller than its expectation value. Our false conjecture claims is that in states which violate $\langle T^s_{\mu\nu} K^\mu K^\nu \rangle_\psi \geq 0$ the fluctuations of the stress tensor are comparable or bigger than its mean. 

In the above $ T^s_{kk}(x)$ is the smeared, regulated, null-null contracted stress-energy operator. 
We start by defining $\langle T_{\mu\nu} \rangle_{\psi}$ as the regulated (vacuum-subtracted) quantity: 
\begin{equation}
\label{regulated SET}
 T_{\mu\nu}  =  \tilde{T}_{\mu\nu}  - \langle 0 \lvert  T_{\mu\nu} \rvert 0 \rangle~,
\end{equation}
where $\tilde{T}_{\mu\nu} $ is the \textit{bare} stress-tensor.
We define a null-contracted smeared version of the above operator. Pick a point $x^\alpha_0$ and a null geodesic $x^\alpha(\lambda)$ passing through this point, with $x^\alpha(0) = x^\alpha_0$. In addition, pick a positive smearing function $f(\lambda)$ centered at $\lambda=0$ with 
\begin{equation}
\int^{\infty}_{-\infty} f(\lambda) d \lambda = 1 ~.
\end{equation}
Then we define the smeared stress tensor by
\begin{equation}
\label{smeared SET}
T^{s}_{kk}(x_0^\alpha) \equiv \int^{\infty}_{-\infty} f (\lambda) T_{\mu \nu}(x^\alpha(\lambda)) \frac{d x^\mu}  {d \lambda} \frac{d x^\nu}  {d \lambda} d\lambda ~.
\end{equation}

\textbf{Strength of Conjecture}: If (\ref{conjecture}) were true, then in the SCG regime all the GR results implied by NEC would automatically hold and the existence of all space-times requiring exotic matter would automatically be forbidden. This would include recent constructions of traversible wormholes in asymptotically Anti-de Sitter spacetime \cite{Gao:2016bin} \cite{Maldacena:2017axo} as well as more exotic wormholes such as those constructed by Visser \cite{Visser:1989kh}.

\subsection{Quantitative measure of fluctuations}
\label{fluctuations}


In order to test our conjecture, we must  distinguish between semiclassical and non-semiclassical states. 
In principle the full probability distribution for the smeared stress tensor is necessary to diagnose whether it is a good approximation to replace it by its mean, but the simplest and most efficient quantity for this purpose is the variance 
\begin{equation}
\sigma^2 = \langle T^s_{kk} (x) T^s_{kk} (x) \rangle_\psi - \langle T^s_{kk} (x) \rangle_\psi \langle T^s_{kk}(x) \rangle_\psi
\end{equation}
 For semiclassical states, the standard deviation $\sigma$ should be much smaller than the mean. On the other hand, for very quantum states, the fluctuations are always close to (or in most cases much bigger) than the mean. 
 
 We should note that the variance should not be used as panacea. Since we are discussing the analysis of a PDF we do not really know, there could be features of the PDF to which the variance is blind. For example, the study of heavy tails, or the asymptotic behavior of the PDF in general, would require the computation of higher cumulants. Nevertheless, as we shall see the variance is already sensitive enough for our purposes.
 
 Note that according to this definition the vacuum is highly quantum. The point here is that if we want to calculate the backreaction on the geometry using the classical Einstein equation, it must be a good approximation to replace the stress tensor by its expectation value. In considering fluctuations around the vacuum, we can neglect backreaction, or treat both the metric and matter fluctuations quantum mechanically, as we discuss later in section 4.

 It is important to clarify that our statement is not that all very quantum states are NEC violating. All we are saying is that the subsets of the Hilbert space  of QFT which corresponds to SCG states and NEC violating states are disjoint.

\begin{figure}[h]
\label{SCG diagram}
\centering
\includegraphics[scale=0.6]{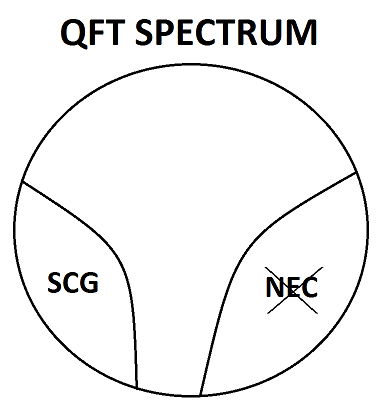}
\caption{This figure provides a ``set theory'' interpretation of our conjecture. Our statement in this context is simply that \textit{the sets of SCG state and NEC violating states are disjoint}. The big circle, represents the full set of QFT states (including Quantum Gravity). This figure, illustrates that NEC violating states are only part of the set of very quantum states. However, as we shall see in \ref{large N} this picture changes for a large number of species.}
\end{figure}

Building on our aforementioned intuition, we introduce the following measure of semiclassicality:

\begin{equation}
\label{Measure of Fluct's rev}
\boxed{\frac{\sigma^2}{\langle T^s_{kk}(x) \rangle^2_\psi} =  \frac{ \langle T^s_{kk} (x) T^s_{kk} (x) \rangle_\psi - \langle T^s_{kk}(x) \rangle^2_\psi} {\langle T^s_{kk}(x) \rangle^2_\psi}}~.
\end{equation}

When $\frac{\sigma^2}{\langle T^s_{kk}(x) \rangle^2_\psi} \ll 1$ the state $\psi$ is semiclassical. When $\frac{\sigma^2}{\langle T_{kk}(x) \rangle^2_\psi} \approx1 ~\mbox{or} > 1$, the state $\psi$ is very quantum. A similar measure of fluctuations was originally introduced by \cite{Kuo:1993if}. However, in that reference normal ordering was used to eliminate certain terms; we believe these terms contain important physical information about the fluctuations of the smeared stress tensor, as pointed out already in \cite{Ford:2000vm}.







\subsection{Testing our conjecture}
\label{tests}

In this section, we present our results for the fluctuations of two representative classes of states.  First, we compute our measure of semiclassicality for the NEC violating superposition of the vacuum and a 2-particle state. We find that the quantity of interest \eqref{Measure of Fluct's rev} is again $\gg 1$; a result which supports our conjecture (\ref{conjecture}). Next, we calculate \eqref{Measure of Fluct's rev} for a class of coherent states, and we show that it is $\ll 1$. Coherent states are considered well-behaved semiclassical states and do not violate NEC. Hence, this last result tests our definition of semiclassicality and provides additional evidence supporting the validity of our conjecture \ref{conjecture}. 

All calculations in this section have been performed in (1+1)-D for massless scalar field, but can be generalized to higher dimensions. The smearing function we used here, is the following simple Gaussian function:

\begin{equation}
\label{smearing f++}
f_{x^+_0}(x^+)= \left(\frac{1}{4\pi\tau^2} \right)^{\frac{1}{2}} e^{-\frac{(x^+_0 -x^+)^2}{2\tau^2}}.
\end{equation}

 $\textbf{x}^+_0$ is the vector which defines the position of the center of the Gaussian, and  $\tau$ determines its width. Let us note that although we picked a specific function, our results hold for \textit{all} choices of smearing functions satisfying the conditions mentioned in \ref{conjecture}.


\medskip

\textbf{Vacuum plus two particles} These states, first introduced by L.H.Ford \cite{Kuo:1993if}, are superpositions of two eigenstates of the number operator. 
Here we choose one specific NEC violating combination of coefficients to minimize the number of tunable parameters, but similar results hold for a wide class of states. Consider
\begin{equation}
\label{0+2 state}
 \lvert \psi \rangle = \frac{\sqrt{3}}{2} \lvert 0_k \rangle + \frac{1}{2} \lvert 2_k \rangle .
\end{equation}
We work in the continuum limit, with commutation relations: 
\begin{equation}
\label{commutation 0+2}
[a_{k},a^{\dagger}_{k'}]  =  \delta (k-k') ~,
\end{equation}
and normalization: 
\begin{equation}
\label{normalization}
\lvert 2_k \rangle = \frac{a^{\dagger}_k a^{\dagger}_k}{\sqrt[]{2}~ L} \lvert 0_k \rangle .
\end{equation}
where L is the size of our system which will appear in our calculation as a $\delta(0)$ IR divergence. For the state (\ref{0+2 state}), the expectation value of our smeared stress energy tensor is:
\begin{equation}
\label{0+2 smeared stress}
\begin{split}
\langle  T^s_{++}(x^+_0) \rangle_{0+2}=\frac{k}{L} \left(1-\sqrt[]{\frac{3}{2}} \cos(2x^+_0 \cdot k) e^{-4\tau^2k^2} \right).
\end{split}
\end{equation}

An important feature of this result is that the negativity of the expectation value emerges from an oscillatory term. If we wish to compute the total energy, we should integrate this quantity over the whole space. That integral of the oscillatory term vanishes, hence the result we will obtain will be strictly positive. The oscillatory behavior suggests that the NEC violating behavior of these kinds of states is connected to interference.

This result is similar to that of \cite{Kuo:1993if}, but due to the smearing the negative contribution on the r.h.s of \eqref{0+2 smeared stress} is exponentially damped. Since the width $\tau$ controls the area of spacetime over which the stress tensor is smeared, \eqref{0+2 smeared stress} has an interpretation familiar from the Quantum Inequalities. The bigger the spacetime region we are considering, the smaller the NEC violating flux.    

The fact that we are in the continuum limit translates to $k L \gg 1$. Taking that into account, one can see that the measure of fluctuations (\ref{Measure of Fluct's rev}) is $\gg1$. But since the full expression is difficult to read, we take the interesting limit $ \frac{1}{\tau} \rightarrow  k $, yielding the following simple result: 
\begin{equation}
\label{0+2 flucts final}
\begin{split}
\frac{\sigma^2_{0+2}}{\langle  T^s_{++} \rangle^2_{0+2}}  \approx \frac{0.0625~ k^2 L^2}{(1-\sqrt[]{\frac{3}{2}} \cos(2x^+_0 \cdot k) e^{-2} )^2}  + \mathcal{O}(k~L) \gg 1. \\
\end{split}
\end{equation}
Result \eqref{0+2 flucts final} clearly shows that the 0+2 particle states are not semiclassical. A more detailed analysis can be found in \cite{Krommydas:2018fgs}.

\bigskip

\textbf{Coherent States.} It is crucial that we also present an example for which the measure of fluctuations \eqref{Measure of Fluct's rev} becomes small. According to our conjecture, this should be a semiclassical state, and as we have discussed the most natural candidates are coherent states. To make our life easier we consider a state which is a superposition of many particles of a single mode. Following Glauber \cite{Glauber}  we use the displacement operator $D_k(z)$ acting on the vacuum to create the states mentioned above.

\begin{equation}
\label{coherent}
\lvert \psi \rangle_{coh} \equiv D_k(z) \lvert 0 \rangle .
\end{equation}
where 
\begin{equation}
\label{dis operator}
D_k(z) \equiv e^{ (z~a^{\dagger}_k -z^{*}~a_k) } = e^{-\frac{\lvert z \rvert^2}{2}k \delta(0)} ~ e^{z~a^{\dagger}_k} ~ e^{-z^{*}~a_k} ,
\end{equation}
and $z= s~e^{i\gamma}$. In \eqref{dis operator} we made use of the Baker-Campbell-Hausdorff formula to obtain the commutation relations we need. Moreover, from this final form of $D_k(z)$ in \eqref{dis operator}, one can immediately observe that the state we obtained is indeed a 1-mode excited coherent state since the action of $e^{-z^{*}~a_k}$ on the vacuum reduces to that of the identity. 

The commutation relations we will need are the following:
\begin{equation}
\label{comm coh}
\begin{split}
[a_{k},a^{\dagger}_{k'}] & =  \delta (k-k')~k\;,  \\
[a_{k},D_{k'}] & =  z~ D_{k'}~ \delta (k-k')~k \;, \\
[D^{\dagger}_{k'},a^{\dagger}_{k}] & = z^{*}~ D^{\dagger}_{k'}~ \delta (k-k')~k  \;. \\ 
\end{split}
\end{equation}

After performing similar calculations to the above two cases, we obtain the following expressions for the expectation value of the stress tensor and its fluctuations: 
\begin{equation}
\label{stress coh}
\begin{split}
\langle T^{s}_{++} \rangle_{coh}=2~k^2~s^2 \left[1-\cos2(\gamma -x^+_0~k) ~ e^{-4\tau^2k^2} \right]\; ,
\end{split}
\end{equation}
\begin{equation}
\label{fluct coh}
\begin{split}
\langle T^{s}_{++} T^{s}_{++} \rangle_{coh}  = & \frac{0.042}{\tau^4} + 2~k^4~ s^4~e^{-8\tau^2k^2} + 2~k^4~ s^4~e^{-8\tau^2k^2} \cos(4\gamma-x^+_0 k) \\ & + 4S^4 k^4 - 8S^4 k^4 \cos2(\gamma-x^+_0 k) e^{-4\tau^2k^2} - \\ & -
\frac{2~k^2~ s^2~e^{-2\tau^2k^2}}{\tau^2}\cos2(\gamma-x^+_0 k)+ 2s^2 \frac{k^2}{\tau^2} e^{-2\tau^2k^2}+ \\ & + \frac{2 k^3~ s^2 ~e^{-2\tau^2k^2}}{\tau}~\sqrt[]{\frac{\pi}{2}}~\left(1+\mbox{Erf} \left(\sqrt[]{2}~k~\tau \right)-\mbox{Erfc}\left(\sqrt[]{2}~k~\tau \right)\right)\;. \\
\end{split}
\end{equation}

It is worth noticing that the quantity $\langle T^{s}_{++} \rangle_{coh}$ is always positive \eqref{stress coh}, and thus the semiclassical coherent states do not violate the quantum version of NEC. All that is left is to compute the measure of the stress tensor fluctuations, and show that it is indeed $\ll 1$.

\begin{equation}
\label{coh flucts}
\begin{split}
\frac{\sigma^2_{coh}}{\langle  T^s_{++} \rangle^2_{coh}} & =  \frac{ \langle T^{s}_{++} T^{s}_{++} \rangle_{coh} -\langle T^{s}_{++} \rangle^2_{coh}}{\langle T^{s}_{++} \rangle^2_{coh}} = \\ & = \Big( \frac{0.042}{\tau^4} + 2~k^4~ s^4~e^{-8\tau^2k^2} + 2~k^4~ s^4~e^{-8\tau^2k^2} \cos 4 (\gamma-x^+_0 k) \\ & + 4s^4 k^4 - 8s^4 k^4\cos2(\gamma-x^+_0 k) e^{-4\tau^2k^2} - \\ & -
\frac{2~k^2~ s^2~e^{-2\tau^2k^2}}{\tau^2}\cos2(\gamma-x^+_0 k)+ 2s^2 \frac{k^2}{\tau^2} e^{-2\tau^2k^2}+ \\ & + \frac{2k^3~ s^2 ~e^{-2\tau^2k^2}}{\tau}~\sqrt[]{\frac{\pi}{2}}~\left(1+\mbox{Erf} \left(\sqrt[]{2}~k~\tau \right)-\mbox{Erfc}\left(\sqrt[]{2}~k~\tau \right)\right)- \\ & - 4~k^4~s^4 \left[1-\cos2(\gamma -x^+_0~k) ~ e^{-4\tau^2k^2} \right]^2 \Big)  / \langle T^{s}_{++} \rangle^2_{coh}\;. \\ 
\end{split}
\end{equation}


It is easy to see from \eqref{dis operator} that $s$ is connected to the number of single mode excitations of our state, and hence to obtain a coherent state $s$ should be big. 
For $s=0$, one gets back the vacuum and the more $s$ grows, the more coherent the state becomes. Even without taking this into account, the exponential damping terms may be enough to make these fluctuations small. Nevertheless, one should also check the interesting limit where these damping terms reduce to $\mathcal{O}(1)$ factors; which is again the limit $ \frac{1}{\tau} \rightarrow k  $. Remembering that $s$ is big, and noticing that it appears in the denominator of \eqref{coh flucts} as $\mathcal{O}(s^4)$ we keep only terms of $\mathcal{O}(s^4)$ in the numerator, after having taken the limit of interest. After taking into account the above considerations, we obtain the following form for our measure of fluctuations in a coherent state:

\begin{equation}
\label{coh flucts limit}
\begin{split}
\frac{\sigma^2_{coh}}{\langle  T^s_{++} \rangle^2_{coh}} & = \Big( 2~k^4~ s^4~e^{-8} + 2~k^4~ s^4~e^{-8} \cos 4 (\gamma-x^+_0 k) \\ & + 4s^4 k^4- 8s^4 k^4\cos2(\gamma-x^+_0 k) e^{-4} - \\ &  - 4~k^4~s^4 \left[1-\cos2(\gamma -x^+_0~k) ~ e^{-4} \right]^2 \Big) / \langle T^{s}_{++} \rangle^2_{coh} = \\ & = 0 (\ll1) \;.\\ 
\end{split}
\end{equation}

Hence, even in this extreme limit, the fluctuations are negligible. 
The above result provides us with evidence to support our claim that for semiclassical states the fluctuations of the stress tensor are small, and subsequently also support our conjecture \ref{conjecture}.

\subsection{Large Number of Species}
\label{large N}

As we have already mentioned our conjecture does not work for field theories with a large number of species N. More precisely, the states we excluded from the SCG spectrum using condition \ref{Measure of Fluct's rev}, satisfy it in the large N limit. Intuitively, this is because the contributions of each field to the fluctuations add up incoherently, but the contributions to the stress tensor add coherently. Quantitatively, if we have $N$ noninteracting species, 
\begin{equation}
\langle T_{kk}^s \rangle_N = N \langle T_{kk}^s \rangle_1 ~,
\end{equation}
while
\begin{equation}
(\sigma^2)_N = N (\sigma^2)_1 ~,
\end{equation}
so that the measure of classicality scales as
\begin{equation}
\label{vanishing flucts}
\frac{\sigma^2}{\langle T^s_{kk}(x) \rangle^2_\psi} \sim \frac{1}{N} ~.
\end{equation}
Therefore, any NEC violating state can be made semiclassical as long as one has the freedom to make a large number of non-interacting copies of the original theory.

We have investigated whether something else goes wrong as the number of fields increases that would invalidate the semiclassical approximation.  One of our attempts  was inspired by I-Sheng Yang's recent work \cite{Yang:2017xyh} \cite{Yang:2017xyh} \cite{Baker:2017sgx}. In this series of papers Yang argues that in a semiclassical theory, the quantum degrees of freedom become significantly entangled with the ``classical'' background degrees of freedom leading to loss of unitarity in both systems. (Unitarity is preserved in the full system.) This becomes a problem for the quantum theory, when the relevant timescale of the problem is comparable to the entanglement timescale, i.e. the timescale at which the two systems become significantly entangled. 

Unfortunately, when we tried to check if such a loss of unitarity occurs faster than the relative timescale in a simple quantum mechanical analog of our large N, NEC violating QFT states, we found that the answer was negative.

It may be the case that our conjecture can be salvaged somehow by modifying it to avoid the problems we have discussed in this section. Regarding the large number of species, one might simply argue that a (very) large number of species does not exist in nature. 



\section{A New Bound}
\label{New Bound}

As we saw in section \ref{large N}, our conjecture (\ref{conjecture}) is violated in theories with a large number of species. Since there is no conclusive evidence against the existence of such theories, the next logical step is to see how large these violations can become. As we have already mentioned \ref{global and local bounds}, local bounds do not place any real restrictions. Interestingly, known semi-local bounds like QI's also fail to bound these states, since the number of species now appears on the l.h.s. leading to their violation. One could imagine solving this issue by an \textit{ad hoc} insertion of an N on the r.h.s., but as we shall discuss on \ref{explicit eg} QI's have further issues. Surprisingly, there are no null-integrated semi-local bounds in the current literature that hold in dimensions higher than 1+1 D.

In the first part of this section \ref{the bound}, we introduce a new bound which solves both problems simultaneously. It fills the gap in the literature for a semi-local bound that holds in any dimension, and restricts the amount of negative null energy density for theories with a large number of species in a natural way. In fact, this bound is much more general than the one we assumed in our false conjecture ( i.e. $\langle T^s_{kk} \rangle_{\psi} \geq 0$ ) since it extends beyond the SCG regime. The generality of this bound enables us to discuss a number of interesting cases which lie outside the SCG spectrum, like the counterpart of Hawking radiation\footnote{Usually, this state is regarded as semiclassical but by our definitions it is just outside this regime, since the fluctuations of the corresponding stress tensor are quite large \cite{Wu:1999ea}. }.

In \ref{derivation/assumption}, we show that bound \eqref{Bound} leads to a number of interesting results. Finally in \ref{explicit eg}, we show how it bypasses all the issues of QI's by providing an analytic calculation for the $0+2$-particle state. To our knowledge, there are no counter examples to this bound, but whether or not a rigorous proof exists is left for future research. 

\subsection{Proposed Bound}
\label{the bound}

As we mentioned, the following bound is much more general than the one appearing in our conjecture (\ref{conjecture}). In order to make clear what its regime of validity is, we formulate it in the form of a proposal:

\paragraph{Proposed Bound.} {\bf For all states within the regime of Perturbative Quantum Gravity (PQG), in any dimension d and for locally flat regions, the expectation value of the null-null contracted Stress-Energy tensor, smeared over any achronal null geodesic is bounded by 

\begin{equation}
\label{Bound}
\boxed{  \langle T^s_{kk} \rangle  \geq - \frac{ B}{G_{N}\tau^2 }~ .} 
\end{equation}
 }


\medskip
In \eqref{Bound} $B$ is a constant we have yet to determine and $\tau$ is the affine distance over which we are smearing; our smearing length, defined more precisely below. The smallest distance we allow ourselves to smear is $l_{uv}$, and the largest is such that we remain in a locally flat region.

\textcolor{blue}{}


The above bound is intuitively clear, but the smearing length is not yet well-defined. Following \cite{Flanagan:1997gn} we make the smearing length precise by proposing the bound 
\begin{equation}
\label{general bound}
\boxed{ \int^{+\infty}_{-\infty} d\lambda f(\lambda) \langle T_{kk} \rangle (\lambda) \geq - \frac{B}{G_N} \int^{+\infty}_{-\infty}  d\lambda \frac{(f'(\lambda))^2}{f(\lambda)} ~.}
\end{equation}
We propose that \eqref{general bound} holds for arbitrary smearing functions $f$ which are smooth and non-negative. The precise bound can be thought of as a definition of the smearing length $\tau$ appearing in the intuitive bound. Explicitly, the bounds are the same if we define the smearing length $\tau$ by
\begin{equation}
{1 \over \tau^2} \equiv \int d\lambda \frac{(f'(\lambda))^2}{f(\lambda)} ~.
\end{equation}
where we have assumed the smearing function $f$ is normalized.
As we shall see in more detail in \ref{explicit eg}, our bound is invariant under affine re-scalings. 

Our bound has the following clear interpretation: the null energy density within a region controlled by the width $\tau$ of our smearing function is bounded $- \frac{ B}{G_{N}\tau^2 }$ . The bound probably does not hold for discontinuous functions $f$. For example, one can trap arbitrarily large negative energy densities at the sharp edges of Theta functions \cite{Flanagan:1997gn}. However, physical apparatuses are restricted by the uncertainty principle and are therefore unable to perform such instantaneous sharp switches \cite{Fewster:2012yh}. 

\textit{Remark}: In general QFT's one is allowed to consider effects of non-local sources. However, the resulting stress-energy operator is not conserved $\nabla^{\mu} T_{\mu\nu} \neq 0 $, and therefore cannot be coupled consistently to gravity. On the other hand, theories like ghost condensate \cite{ArkaniHamed:2003uy} will presumably violate our bound, while there is no obvious obstacle to coupling them to gravity. These theories are believed to lie in the swampland \cite{Dubovsky:2006vk}.

\medskip

Our bound (with fixed order one constant $B$)  may hold even for field theories with an arbitrary number of fields. That is because of the presence of the renormalized Newton's constant $G_N$. When gravity is coupled to theories with a large number of species, the renormalized $G_N$ to 1-loop order and in 4-D becomes $G_N \sim \frac{l^2_{UV}}{N}$ \footnote{By large number of fields here, we mean a large number of \textit{scalar} species. For spinors and vector fields the story goes a bit differently, but the relevant results will remain the same.} \cite{Kaloper:2015jcz}. We are not the first to make use of $G_N$ to solve the species problem. The Covariant Entropy bound, or Bousso Bound \cite{Bousso:1999xy}, which is an improvement of the Bekenstein Bound, solves the species problem of the latter in a similar way. Later, Cassini produced another beautiful generalization of the bound \cite{Casini:2008cr}.






\begin{figure}[h!]
\label{PQG spectrum}
\centering
\includegraphics[scale=0.75]{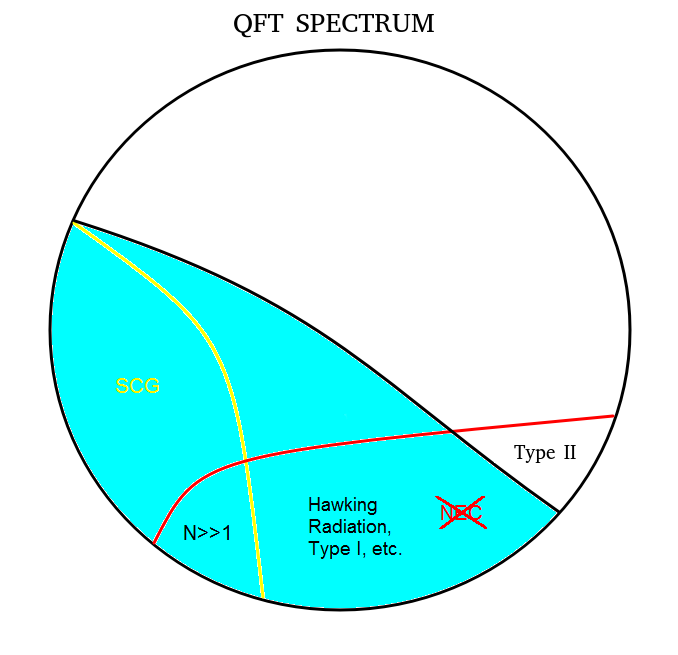}
\caption{The light blue represents all PQG states. Our statement, is that all states in this region satisfy \eqref{Bound}. The region enclosed by the red line corresponds to the set of NEC violating states and the region enclosed by the yellow line to the set of SCG states. Notice that in this picture, there is an overlap between the SCG and NEC violating spectrum. The union of these regions represents NEC violating states in theories with a large number of species. The white part of the diagram represents states which cannot be treated within well-understood theories (for example states with large backreaction). `Type I' refers to NEC violating states with small backreaction, while `Type II' refers to NEC violating states with large backreaction.}
\end{figure}

One of the configurations which prevent the generalization of our bound to larger regions of integration is the accumulated effect of Hawking Radiation. If we smear over a null geodesic close to the horizon of an evaporating Black Hole,
and for a long affine distance (many times the curvature radius), we can amass enough negative energy to violate bound \eqref{Bound}. What is more, this story can be extended to certain Unruh paths using an equivalence principle argument. Hence, we chose to avoid this issue by imposing that our smearing is performed over a region of size comparable to the radius of curvature. We point out here that this accumulative feature of the Hawking radiation goes against most prior intuitions on how negative energy densities are bounded. For example, ANEC and QI's tell us that the bigger the region one integrates over, the less negative energy it can contain.

As in \ref{conjecture}, we provide a qualitative picture of the set of states our bound restricts. In this case the diagram \ref{PQG spectrum} is somewhat more crowded and our bound \eqref{Bound}, although more general, is subject to certain limitations (e.g. for certain spacetimes, the smearing length cannot be arbitrarily large). However, that should not cause any confusion since these limitations are universal, i.e. they are assumed for all states. Keeping these limitations in mind, we can interpret \ref{PQG spectrum} the same way we interpreted \ref{SCG diagram}.

The NEC violating states we excluded in \ref{fluctuations} due to their large fluctuations, are separated in two types. The classification to type I and type II is done based on their back-reaction; significant back-reaction being $\langle T_{kk} \rangle G_{N} \tau^2 \approx 1 $. 
 At the boundary between PQG and type II lie the states which saturate our bound. We have yet to determine which these states are, but we note that identifying them will make for an even more precise version of our bound, and provide evidence for its validity.



\medskip






\bigskip

\subsection{Consequences of the Bound}
 \label{derivation/assumption}


In what follows, the consequences of bound  \eqref{Bound} for a number of physical set-ups are discussed. 



\medskip

\paragraph{Defocusing.} The first scenario in which one can see our bound emerge is the passing of a null congruence through a lump of negative energy. To study this configuration we use the optical (null) Raychaudhuri equation, which in 3+1-d takes the following form: 

\begin{equation}
\label{Raychau}
\frac{d\theta}{d\lambda} = - G_N \langle T_{kk} \rangle + \omega_{\mu\nu} \omega^{\mu\nu} -\sigma_{\mu\nu} \sigma^{\mu\nu}  - \frac{1}{2} \theta^2 ~ .
\end{equation}

For simplicity, we chose a congruence with vanishing rotation $\omega_{\mu\nu}$, shear  $\sigma_{\mu\nu}$ and initial expansion $\theta$.

\begin{figure}[h!]
\centering
\includegraphics[scale=0.75]{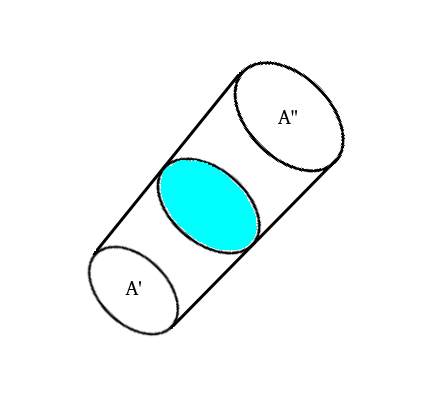}
\caption{Null congruence of initial area $A' $ going through a concentration of negative energy (light blue). The null congruence emerges dilated, but our bound guarantees that the growth of its area remains small.}
\label{defocusing}
\end{figure}

The expansion measures the rate of change of the size of a cross-sectional area element $A$. In terms of $A$, the expansion reads:

\begin{equation}
\label{defoc bound}
\theta =  \lim_{A \to 0} \frac{1}{A} \frac{dA}{d\lambda} ~ .
\end{equation}

Integrating \eqref{Raychau} twice for a generic form of the stress tensor (\ref{defocusing}), and assuming bound \eqref{Bound}, one obtains:
\begin{equation}
\boxed{\frac{\Delta A}{A} = - \lambda^2 G_N \langle T_{kk} \rangle \geq -B ~.}
\end{equation}
Hence, our bound puts a restriction on the amount of ``defocusing'' of a null congruence.


\paragraph{Negative Mass Black Holes.} Another consequence of our bound is that it places a restriction to the magnitude of the gravitational potential of a uniform, spherical, negative energy distribution.

\begin{equation}
\label{grav pot bound}
\boxed{ \frac{G_N M}{r} =  \frac{G_N r^3 \langle T \rangle }{r}  = G_N r^2 \langle T \rangle \geq -B ~,}
\end{equation}


To appreciate the strength of this restriction, we explore this result a little further:

\begin{equation}
\label{zeroth grav pot}
G_N r^2 \langle T \rangle = \frac{G_N N}{r^2}  \approx \frac{-N}{(M_{pl,bare}^2 + N M_{uv}^2 )r^2} \approx - \frac{l_{uv}^2}{r^2} ,
\end{equation}

where $l_{uv}$ is the UV cut-off, and $(M_{pl,bare}^2 + N M_{uv}^2)$ is the renormalized Planck mass to one loop order. This means that for the zeroth order gravitational potential to be significant, the radius of the distribution should be comparable to the UV cut-off.

\subsection{0+2 state; an explicit example}
\label{explicit eg}

In \cite{Fewster:2002ne}, the authors claimed that no null-worldline inequalities exist in 4-D. Their first argument, which was originally introduced by Ford \cite{Ford:1994bj}, was that a null-worldline QI will not be invariant under rescaling of the affine parameter. Furthermore, they showed via a specific example that the expectation value of the null-smeared stess-energy operator is unbounded.
In this section, we show that their claims do not apply to our bound.

For the rescaling argument, the reason is simple. Our bound has the form
\begin{equation}
{1 \over \lambda_2 - \lambda_1} \int_{\lambda_1}^{\lambda_2} d \lambda \langle T^s_{\mu\nu}\left(x^\alpha(\lambda)\right) \rangle_\psi ~{d x^\mu \over d \lambda}{d x^\nu \over d \lambda }  \geq - \frac{B}{G_N\lambda^2}~.
\end{equation}
Under the rescaling $\lambda \rightarrow a \lambda$, both sides scale in the same way. The argument does succeed in eliminating the possibility of quantum-inequality like bounds, which have the general form: 
\begin{equation}
{1 \over \lambda_2 - \lambda_1} \int_{\lambda_1}^{\lambda_2} d \lambda
\langle T^s_{\mu\nu}\left(x^\alpha(\lambda)\right) \rangle_\psi ~{d x^\mu \over d \lambda}{d x^\nu \over d \lambda }
  \geq - \frac{C}{\lambda^4}~,
\end{equation}
where it is clear that their argument applies. 

Now we need to address their claim that the smeared null energy can be arbitrarily negative. In what follows, we calculate the expectation value of the null-smeared stess-energy operator in their explicit example state, with the additional assumption that the momenta are bounded by the UV cutoff. We explain the quantities and assumptions relevant for our result, but for a more detailed understanding we refer the reader to the original work \cite{Fewster:2002ne}.

To be clear, the calculation of \cite{Fewster:2002ne} is correct, but we are imposing an additional rule that the momenta involved in making the state must lie below the UV cutoff of the theory. We will find that once this rule is imposed our bound is respected.

Their basic idea is to take a NEC violating state and perform a large boost to enhance the NEC violation in a given null interval. First, one can write the general $0+2$ particle state in the following way:

\begin{equation}
\label{general state}
\psi_\alpha = N_{\alpha} \left[ \lvert 0 \rangle + \int \frac{d^3\textbf{k}}{(2\pi)^3} \frac{d^3\textbf{k}'}{(2\pi)^3} b_{\alpha}(\textbf{k},\textbf{k}') \lvert \textbf{k},\textbf{k}' \rangle \right] ~,
\end{equation}

so the normalization is:

\begin{equation}
\label{normalization example gen}
N_{\alpha} = \left[ 1 + 2 \int \frac{d^3\textbf{k}}{(2\pi)^3} \frac{d^3\textbf{k}'}{(2\pi)^3} \lvert b_{\alpha}(\textbf{k},\textbf{k}') \rvert^2 \right]^{-\frac{1}{2}}~.
\end{equation}

In general, $b_{\alpha}$ is a complicated function whose purpose is to ensure that the states satisfy certain reasonable conditions. In this example, under a specific choice of $b_{\alpha}$ 
the normalization becomes:

\begin{equation}
\label{normalization example}
N_{\alpha} = \left[ 1 + \frac{\alpha^{2\sigma - 6}}{128\pi^4} \right]^{-\frac{1}{2}}~,
\end{equation}

and the expectation value of the energy density takes the form:

\begin{equation}
\label{energy density}
\langle \rho (t,0,0,z) \rangle_{\omega_{\alpha}} = \frac{ 2 N^2_{\alpha}}{16 \pi^4 \Lambda^4_0} \mbox{Re} \left( \lvert \xi_{\alpha} (t,z) \rvert^2 \frac{\alpha^{2\sigma - 7}}{24\pi^2} - \xi_{\alpha} (t,z)^2 \alpha^{\sigma - 4}  \right) ~,
\end{equation}

with 

\begin{equation}
\label{xi}
 \xi_{\alpha} (t,z) = \int^{\Lambda_0}_0 dv e^{-iv(t-z)/\alpha} \left[ \frac{iv^2}{z}  e^{-ivz} + \frac{v}{z^2} \left( e^{-ivz} -1 \right)  \right] ~.
\end{equation}

In the above $\sigma = 3.75$. $\frac{\Lambda_0}{\alpha} = \Lambda$ where $\Lambda$ is the momentum cutoff and the limit $\Lambda \rightarrow \infty$ corresponds to $ \alpha \rightarrow 0$, and $\omega_{\alpha}$ just labels different Hadamard states. The quantity $v$ which appears in \eqref{xi}, is $v\equiv k \alpha$ where $k$ are the 3-momenta. Throughout our calculation, we take several approximations which we keep track of and demonstrate their validity at the end. We first compute the integral $\xi_{\alpha} (t,z) $. Our first approximation is to expand the quantity which appears in brackets of \eqref{xi} around $v  \rightarrow 0$. 

\begin{equation}
\label{brachet xi}
 \frac{iv^2}{z}  e^{-ivz} + \frac{v}{z^2} \left( e^{-ivz} -1 \right) = \frac{v^3}{2} - \frac{iv^4z}{3} + \mathcal{O}[v^5]~.
\end{equation}

Taking \eqref{brachet xi} into account and making the substitution $y\equiv v(t-z)/\alpha$, we obtain: 

\begin{equation}
\label{xi with y}
 \xi_{\alpha} (t,z) = \int^{(t-z) \frac{\Lambda_0}{\alpha}}_0 dy~ e^{-iy} \left[ \frac{y^3}{2} \left(\frac{\alpha}{t-z} \right)^4 - \frac{iy^4z}{3} \left(\frac{\alpha}{t-z} \right)^5 \right] ~.
\end{equation}

Although it is not easy to see at this stage, the $O[y^4]$ term gives less divergent contributions to the final results, so we drop it. Our second approximation is that $y\equiv v(t-z)/\alpha << 1$ so that $e^{-iy} \approx 1$ (2). Then the answer is simply: 

\begin{equation}
\label{xi result}
 \xi_{\alpha} (t,z) = \Lambda^4_0 ~,
\end{equation}

The only regime in which the integral of the energy density can become negative is when $\alpha \rightarrow 0$. Although in our case $\alpha$ will not become exactly zero, it will be small enough for the following approximations to hold:

\begin{equation}
\label{normalization example approx}
N_{\alpha} = \left[ 1 + \frac{\alpha^{3/2}}{128\pi^4} \right]^{-\frac{1}{2}} \approx 1~,
\end{equation}

and 

\begin{equation}
\label{energy density}
\langle \rho  \rangle_{\omega_{\alpha}} = \frac{ -1}{8 \pi^4 \Lambda^4_0} \alpha^{-1/4} \left(   \frac{\Lambda^8_0}{ 64}   \right) = - \Lambda^2 \Lambda^2_0 \alpha^{7/4} ~ 8.47 \cdot 10^{-7} = - \frac{\alpha^{7/4} \Lambda^2_0 }{l^2_{uv}} ~ 8.47 \cdot 10^{-7}  ~.
\end{equation}

It is now simple to smear this result over the region of interest (here we choose a region which again is the most likely to cause problems). Following \cite{Fewster:2002ne} we smear along the geodesic $(\lambda,0,0,\lambda)$ where $\lambda$ is an affine parameter, to find:

\begin{equation}
\label{energy density smear}
\Lambda \int^{1/\Lambda}_0 d\lambda \langle \rho \rangle_{\omega_{\alpha}} = - \sqrt[]{2} \frac{\alpha^{7/4} \Lambda^2_0 }{l^2_{uv}} ~ 8.47 \cdot 10^{-7} > - \frac{1}{G_N \lambda^2}~.
\end{equation}

Thus, we see that our bound is satisfied. As promised, we see that in the integral of interest $t=z=\lambda$, so assumption (2) which essentially let us ignore the oscillating term in \eqref{xi with y} is satisfied.

\bigskip

\subsection{The fluctuations of a smeared stress tensor}

The 2-point function of the stress tensor is famously divergent. Even for the simple free massless scalar in the vacuum state, one needs to impose a severe regularization in order to obtain a finite result. As in previous sections, we use smearing as our regularization scheme. The following simple calculations show that not all directions are good enough to cure the divergences of these fluctuations. Namely, the fluctuations of the stress tensor are finite when one smears it along a timelike direction, or along at least both null directions (which is equivalent to smearing over a space and a time direction). All other types of smearing of the stress tensor fail to tame the divergences of its 2-point function.\footnote{This result seems to be known to researchers in the field but we have not found a reference in the literature, so we calculate the result here.}

\medskip

We show this in the simplest example of a free massless field in 3+1 dimensions.  In lightcone coordinates, the field can be expressed as
\begin{equation}
\label{scalar field light cone}
\phi = \int \frac{dk^- d^2k^{\perp}}{k^-} \left( \alpha ~ f(k) + \alpha^{\dagger} ~ f^*(k) \right) ~,
\end{equation}
where 
\begin{equation}
\label{f in light cone}
f(k)= e^{-i(k^- x^+ + k^+ x^- - 2k^{\perp} x^{\perp})}
\end{equation}
and $\alpha$'s have the commutation relations
\begin{equation}
\begin{split}
& [\alpha_k, \alpha^{\dagger}_{k'}] = 2 (\pi)^3 k^- \delta^3 (k-k'), \\ & [\alpha_k, \alpha_{k'}] = 0 , \\& [\alpha^{\dagger}_k, \alpha^{\dagger}_{k'}] = 0.
\end{split}
\end{equation}

The null stress tensor is then:
\begin{equation}
\label{T plus plus}
\begin{split}
T_{++}  = & \int dk_1^- d^2k_1^{\perp} dk_2^- d^2k_2^{\perp}  \cdot ~ (\alpha_1^{\dagger} \alpha_2^{\dagger} ~ e^{i\left[\left(k_1^- + k_2^- \right) x_1^+ + \left(k_1^+ + k_2^+  \right)x^- - 2\left(k_1^{\perp} + k_1^{\perp} \right) x^{\perp}\right]} +\\ & + h.c. + vanishing ) ~.
\end{split}
\end{equation}
By \textit{vanishing} here, we mean all terms which will not contribute to the 2-point function of the stress tensor. If we do not smear along any direction the 2-point function yields the following simple result:
\begin{equation}
\langle T_{++}T_{++} \rangle_0 = \int dk_1^- d^2k_1^{\perp} dk_2^- d^2k_2^{\perp}~ k_1^- k_2^- ~ ,
\end{equation}
which is clearly badly divergent. Smearing over the null direction, using the usual Gaussian \eqref{smearing f++}, one finds that the result is still divergent in the transverse directions.
\begin{equation}
\langle T_{++}T_{++} \rangle_0 = \int dk_1^- d^2k_1^{\perp} dk_2^- d^2k_2^{\perp}~ k_1^- k_2^- ~ e^{-\tau^2\left(k_1^- +k_2^- \right)^2} ~ .
\end{equation}

One might think that the divergences could be regulated by smearing over the transverse directions. However, this is still not enough: 
\begin{equation}
\langle T_{++}T_{++} \rangle_0 = \int dk_1^- d^2k_1^{\perp} dk_2^- d^2k_2^{\perp}~ k_1^- k_2^- ~ e^{-\tau^2\left(k_1^- +k_2^- \right)^2} ~ e^{-2\sigma^2\left(k_1^\perp +k_2^\perp \right)^2} ~ .
\end{equation}
since the transverse divergences are still not suppressed when $k_1^\perp \rightarrow - k_2^\perp $. To make the fluctuations finite, one needs to smear over at least both $x^+$ and $x^-$ directions.

\begin{equation}
\label{T plus plus}
\begin{split}
T_{++}  = & \int dk_1^- d^2k_1^{\perp} dk_2^- d^2k_2^{\perp}  ~  ( \alpha_1^{\dagger} \alpha_2^{\dagger} ~ e^{-\tau^2\left(k_1^- +k_2^- \right)^2} ~  e^{-\rho^2\left(\frac{\left(k_1^{\perp}\right)^2}{k_1^-} +\frac{\left(k_2^{\perp}\right)^2}{k_2^-} \right)^2}  ~ \cdot \\ & \cdot  e^{- 2 i \left(k_1^{\perp} + k_1^{\perp} \right) x^{\perp}} + h.c. ) ~,
\end{split}
\end{equation}
where we have taken into account the dispersion relation of the light-cone momenta, namely:
\begin{equation}
\label{dispersion light cone}
k^+k^-=\left(k^{\perp}\right)^2 ~.
\end{equation}

The two point function of \eqref{T plus plus} in the vacuum is:
\begin{equation}
\langle T_{++}T_{++} \rangle_0 = \int dk_1^- d^2k_1^{\perp} dk_2^- d^2k_2^{\perp}~ k_1^- k_2^- ~ e^{-\tau^2\left(k_1^- +k_2^- \right)^2} ~  e^{-\rho^2\left(\frac{\left(k_1^{\perp}\right)^2}{k_1^-} +\frac{\left(k_2^{\perp}\right)^2}{k_2^-} \right)^2}  ~ .
\end{equation}

Making the substitution
\begin{equation}
x_1 = \frac{\left(  k_1^{\perp} \right)^2}{k_1^-} \rho  \Leftrightarrow dk_1^{\perp} = dx_1 \sqrt[]{\frac{k_1^-}{\rho ~  x_1 }}~,
\end{equation}
The 2-point function yields:
\begin{equation}
\langle T_{++}T_{++} \rangle_0 = \int dk_1^-  dk_2^- ~ \left( \frac{ k_1^- k_2^-}{\rho } \right)^2  ~ e^{-\tau^2\left(k_1^- +k_2^- \right)^2} \approx   \frac{1}{\rho^2 \tau^6} ~,
\end{equation}
which is finite since $k_1^-,k_2^-$ play now the role of positive frequencies. 

\medskip

 Looking at these results, one may be tempted to interpret them using some heuristic Uncertainty Principle argument: the fluctuations of the stress tensor are finite only when one smears over a non-zero proper time interval. Computing these fluctuations without smearing over time would be like trying to measure the fluctuations of the energy in any arbitrarily small time window, which according to $\Delta E ~ \Delta t \geq \frac{\hbar}{2} $ cannot give a finite answer. However, this argument does not explain why smearing along a null direction is sufficient to regulate the divergence in 1+1 dimensions. In that case, the UV theory is a CFT where $T_{++}$ is a function of $x^+$ only, so smearing over time is the same as smearing over the null direction.



\bigskip




\subsection*{Acknowledgements}
We had many interesting and helpful discussions with colleagues over the course of this project, and apologize for forgetting to acknowledge everyone. We would like to thank  Vassilis Anagiannis, Dionysios Anninos,  Raphael Bousso, Larry Ford, Diego Hofman, Stephen Shenker, Douglas Stanford, and I-Sheng Yang for stimulating discussions. We particularly thank Stefan Leichenauer and Ken Olum for contributing important insights.

\newpage

\appendix
\section{Perturbed vacuum states}
\label{Perturbed vacuum states}

Below, we compute the expectation value of the stress tensor for this state, in 1+1-D light-cone coordinates in Minkowski spacetime. The stress tensor for the free massless scalar in light-cone coordinates has the components:


\begin{equation}
\begin{split}
T_{++} & = \partial_{+}\Phi(z)\partial_{+}\Phi(z), \\
T_{--} & = \partial_{-}\Phi(z)\partial_{-}\Phi(z).
\end{split}
\end{equation}

where we have chosen to work in the coincident limit. From now on, we use the notation $\Phi(z) = \Phi_z$ for aesthetic reasons. Next, we compute the expectation value of $T_{++}$ up to order $J^2$, using the expansion:

\begin{equation}
\label{T++ perturbed}
\begin{split}
\langle T_{++} \rangle_{per.} & = \langle 0 \rvert e^{-i\int J_x d^{2}x\Phi_x} T_{++} e^{i\int J_x d^2x\Phi_x}\lvert 0 \rangle \\
& = \langle 0 \rvert \left(1 -i\int d^2x J_{x}\Phi_{x}-\frac{1}{2}\int d^2x d^2w J_{x}J_{w}\Phi_{x}\Phi_{w} \right) \cdot \left( \partial_{+}\Phi_{z}\partial_{+}\Phi_{z} \right) \cdot \\ 
& ~~~~ \cdot \left(1 + i\int d^2w J_{w}\Phi_{w}-\frac{1}{2}\int d^2x d^2w J_{x}J_{w}\Phi_{x}\Phi_{w} \right)  \lvert 0 \rangle.
\end{split}
\end{equation}

\begin{enumerate}
\item Order $J^0$: one obtains the usual VEV, $\langle 0 \lvert T_{++} \rvert 0 \rangle$ which has already been extensively studied. 

\item Order $J^1$: These are 3-point functions, and it is a well-known fact that for a free scalar field all odd n-point functions vanish. A heuristic way to understand this is by considering the Feynman diagram for a 3-point function; this is just the 3-point vertex. Since there are no interactions, there is no mechanism through which a ``particle'' can decay in two or two ``particles'' can combine in one. A slower, more rigorous way to arrive at this conclusion is to turn to the path integral formalism and obtain the n-point functions. 

\item Order $J^2$: These terms do contribute to \eqref{T++ perturbed} and we shall devote the rest of this section in calculating their contribution. 
\end{enumerate}

There are three terms of order $J^2$. Our first step is to deal with their cumbersome 4-point functions, by invoking the \textit{Wick's Theorem}:

\begin{equation}
\begin{split}
A & =  \langle   \Phi_{x}\partial_{+}\Phi_z \partial_{+}\Phi_z  \Phi_{w}   \rangle_0 \\
  & = 2 \cdot \langle   \Phi_{x} \partial_{+}\Phi_z \rangle_0 \cdot \langle   \partial_{+}\Phi_z \Phi_{w} \rangle_0\;,
\end{split}
\end{equation}

\begin{equation}
\begin{split}
B & = - \frac{1}{2}\langle   \Phi_{x}\Phi_{w} \partial_{+}\Phi_z\partial_{+}\Phi_z \rangle_0 \\
  & = - \langle   \Phi_{x} \partial_{+}\Phi_z  \rangle_0 \cdot  \langle   \Phi_{w} \partial_{+}\Phi_z \rangle_0\;,
\end{split}
\end{equation}

\begin{equation}
\begin{split}
C & = - \frac{1}{2}\langle  \partial_{+}\Phi_z\partial_{+}\Phi_z \Phi_{x}\Phi_{w}  \rangle_0 \\
  & = - \langle \partial_{+}\Phi_z  \Phi_{x}   \rangle_0 \cdot  \langle  \partial_{+}\Phi_z \Phi_{w}  \rangle_0\;.
\end{split}
\end{equation}

By simply adding A,B and C one obtains:

\begin{equation}
\begin{split}
\label{A+B+C}
A+B+C & = \langle \partial_{+}\Phi_z  \Phi_{w}   \rangle_0 \cdot \left(   \langle   \Phi_{x} \partial_{+}\Phi_z \rangle_0 - \langle \partial_{+}\Phi_z  \Phi_{x}   \rangle_0 \right) + \\
& +  \langle \Phi_{x} \partial_{+}\Phi_z  \rangle_0 \cdot \left(   \langle \partial_{+}\Phi_z  \Phi_{w}   \rangle_0 - \langle   \Phi_{w} \partial_{+}\Phi_z \rangle_0  \right).
\end{split}
\end{equation}

Since all operators are \textit{Hermitian}, using the identity $\langle AB \rangle =\langle BA \rangle^*$ and $\langle AB \rangle - \langle AB \rangle^*=2i \cdot \mbox{Im}\langle AB \rangle$ for \eqref{A+B+C} , we arrive to:

\begin{equation}
\begin{split}
\label{A+B+C 2}
A+B+C & = 2i \cdot \langle \partial_{+}\Phi_z  \Phi_{w}   \rangle_0 \cdot  \mbox{Im}  \langle   \Phi_{x} \partial_{+}\Phi_z \rangle_0  + \\
& + 2i \cdot  \langle \Phi_{x} \partial_{+}\Phi_z  \rangle_0 \cdot  \mbox{Im} \langle \partial_{+}\Phi_z  \Phi_{w}   \rangle_0 \;.
\end{split}
\end{equation}

The Feynman propagator in two dimensions reads: 

\begin{equation}
\label{Feynman prop}
\langle \Phi_{x} \Phi_z \rangle_0 = - \frac{1}{4\pi} \ln \frac{1}{(z_+ - x_+)(z_- -x_-)- i\epsilon } \;.
\end{equation}


Remembering that all the derivatives are $\partial_+ = \frac{\partial}{\partial z^+}$, we use \eqref{Feynman prop} to compute all the relevant quantities in \eqref{A+B+C 2}:

\begin{equation}
  \langle   \Phi_{x} \partial_{+}\Phi_z \rangle_0  = \frac{\partial}{\partial z^+} \langle  \Phi_{x} \Phi_z \rangle_0 =  \frac{1}{4\pi}  \frac{1}{(z_+ - x_+)- i\epsilon }\;,
\end{equation}

\begin{equation}
\label{Im 1}
 \mbox{Im} \langle   \Phi_{x} \partial_{+}\Phi_z \rangle_0  =  \frac{\epsilon}{4\pi}  \frac{1}{(z_+ - x_+)^2- \epsilon^2 }\;,
\end{equation}

\begin{equation}
  \langle  \partial_{+}\Phi_z \Phi_w \rangle_0  = - \frac{\partial}{\partial z^+} \langle  \Phi_z \Phi_w \rangle_0 = - \frac{1}{4\pi} \frac{1}{(w_+ - z_+)- i\epsilon }\;,
\end{equation}

\begin{equation}
\label{Im 2}
 \mbox{Im}  \langle  \partial_{+}\Phi_z \Phi_w \rangle_0  =  \frac{\epsilon}{4\pi} \frac{1}{(w_+ - z_+)^2- \epsilon^2 }\;.
\end{equation}


In the epsilon prescription we have used, there is always an implicit limit of $\epsilon \rightarrow 0$. Taking this limit we see that \eqref{Im 1} and \eqref{Im 2} are nothing more than an alternative definition for the delta function:

\begin{equation}
\label{delta func}
\delta(x)= \frac{1}{\pi} \lim_{\epsilon\to0} \frac{\epsilon}{x^2 + \epsilon^2}\;.
\end{equation}

After having brought the 4-point functions to the form we wanted, computing the integrals for the order $J^2$ terms becomes an easy task:

\begin{equation}
\label{Pert result}
\begin{split}
\mathcal{O}(J^2) = & -\frac{2i}{16\pi}\int dx_+ dx_- \int dw_+ dw_- J(x_+,x_-) J(w_+,w_-) \frac{1}{(w_+ - z_+)- i\epsilon } \delta (z_+-x_+) - \\
& -\frac{2i}{16\pi}\int dx_+ dx_- \int dw_+ dw_- J(x_+,x_-) J(w_+,w_-) \frac{1}{(z_+ - x_+)- i\epsilon } \delta (w_+-z_+) = \\
& = -\frac{2i}{16\pi}\int dx_- \int dw_+ dw_- J(z_+,x_-) J(w_+,w_-) \frac{1}{(w_+ - z_+)- i\epsilon } - \\
& -\frac{2i}{16\pi}\int dx_+ dx_- \int dw_- J(x_+,x_-) J(z_+,w_-) \frac{1}{(z_+ - x_+)- i\epsilon } \;.
\end{split}
\end{equation}

Relabeling the last two integrals we obtain:

\begin{equation}
\label{Pert result 2}
\begin{split}
\mathcal{O}(J^2) & =  -\frac{i}{8\pi}\int dx_+ dx_- \int dw_- J(x_+,x_-) J(z_+,w_-) \left( \frac{1}{(z_+ - x_+)- i\epsilon } - \frac{1}{(z_+ - x_+) + i\epsilon } \right) = \\ &  = -\frac{i}{8\pi}\int dx_+ dx_- \int dw_- J(x_+,x_-) J(z_+,w_-)  \frac{2i \epsilon }{(z_+ - x_+)^2 + \epsilon^2 }  = \\ &  = \frac{1}{4\pi}\int dx_+ dx_- \int dw_- J(x_+,x_-) J(z_+,w_-) \delta (x_+ - z_+) = \\ &  = \frac{1}{4\pi}   \left( \int dx_-  J(z_+,x_-)  \right)^2 \;.
\end{split}
\end{equation}


As we mentioned, J is an arbitrary function which describes the form of the source distribution. Finally, the result for the entire expectation value of the stress tensor up to second order in J is:

\begin{equation}
\langle T_{++} \rangle_{per.} = \langle T_{++} \rangle_0 + \frac{1}{4\pi}   \left( \int dx_-  J(z_+,x_-)  \right)^2.
\end{equation}

\end{document}